# Potential Negative Impact on Reliability of Distributed Generation under Temporary Faults

Xuanchang Ran, Tianqi Hong

Department of Electrical and Computer Engineering, New York University, Five Metro-tech Center, Brooklyn, NY, US.

**Abstract:** This paper uncovers potential negative impacts on the SAIFI reliability index produced by the installation of Distributed Generation (DG) in distribution networks when subjected to temporary faults. Detailed network modelling produces accurate time-domain simulations which show how the negative effects on reliability index occur. The problem is caused by fuse opening due to the wrong location of the DG and unsuitable relay settings. A 4 kV urban distribution system is chosen for the analysis. Over 50% of the possible locations where DG can be installed in the network under study yield reduce reliability.

## 1. Introduction

Distributed generation (DG) has become increasingly popular since environment protection and energy usage efficiency have evolved as some of the world's most important concerns. DG is definitely beneficial, but also brings issues to power system stability and power quality which need to be solved [1].

Electric power system reliability indexes are important measures of the overall situation of a power grid and are used by power utilities for management, maintenance, and direct investments. System Average Interruption Duration Index (SAIDI), Customer Average Interruption Duration Index (CAIDI) and System Average Interruption Frequency Index (SAIFI) are the ones of the most valued indexes. Utilities usually evaluate the indexes for each feeder and sub-network [2].

A drop in the indexes means that a network is safer, an increase beyond certain value means that the utility needs to spend money reinforcing the system. Typical action include: installation of faster acting fuses, application of better power dispatch strategies, or renewing aged cables and apparatuses.

Many types of DG exist in the power system such as: synchronous generators, induction generators, PVs, fuel cells, etc. This research focuses on the impact analysis that synchronous generators have on SAIFI in an existing urban 4 kV loop network (see Fig. 1).

The modern urban electric power system is a complex network with varied configurations. The primary feeders are mostly radial. Their reliability calculation is straightforward. In contrast, the low voltage secondary networks are often built with a mesh structure. Thus the evaluation of its reliability seems unnecessary as each load is supplied by multiple paths which introduce extremely high reliability. Therefore, this study focuses solely on the 4 kV loop distribution network which is composed of multiple branches and various kinds of protective devices (Fig. 1).

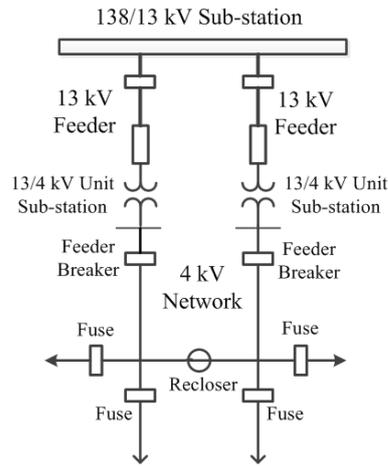

*Fig. 1.* *Configuration of a typical 4 kV loop distribution network*

The effects of DG on the distribution system have been extensively studied in recent years. In [3], the positive effects of distributed generation as a backup system are studied. The effects of different penetration levels at different locations in the distribution system are illustrated. Reference [4] presents an investigation of the impact of DG penetration on the reliability of a radial distribution network. The reliability index is assessed with analytical methods and Monte Carlo simulations. It is shown in [4] how to select the best place to install DG in a radial system. In [5], an analytical approach of reliability assessment for wind-diesel hybrid system with battery banks is proposed. In [6], dynamic disturbances caused by the presence and methods to improve power quality and reliability using low power level DGs were discussed. In [7] a methodology is proposed to allocate DG within certain constraints to minimize losses, reducing the interruption costs, and correcting voltage profile. In [8] the methods to optimize and coordinate the placement of DGs and re-closers with distribution network constraints are studied.

Despite the fact that reliability can be generally boosted when adding DG, the possibility that distributed generators can produce negative effects should not be neglected.

There is always a tendency to install DG in heavily populated areas to share the loads with the power grid. This is so because the closer the DG is to the loads, the lower the power loss dissipated on the cables becomes. It has often been seen that a generator is installed in a feeder branch with heavy loading. If the branch happens to be located downstream of a low rating fuse or a fast acting protective device, the presence of DG could results in undesirable consequences: the contribution of the DG to the fault current could disconnect customers by tripping a device.

This paper show how SAIFI can be increased, in certain cases, when switching devices clear temporary faults. Only the reliability variation in steady state is considered in the calculation of SAIFI, since utilities exclude the calculation of temporary disturbances lasting seconds as they do not cause long term power loss. SAIFI is computed as:

$$\text{SAIFI} = \frac{\text{total number of customers interrupted}}{\text{total number of customers served}}$$

Different from other analyses in the literature, the present study focuses on time-domain dynamic short circuit simulations of a peculiar, yet currently available in the field, 4 kV loop distribution system configuration with every component modelled in detail.

## 2. System and relay modelling

Time domain (transient) simulations need detailed models of al system components, including: voltage source, substations, shunt capacitors, transformers, switches, controllers, regulators, feeder sections, and loads. All components are modelled in this paper as three-phase balanced. The modelling and time domain simulation of a distribution system as the one in Fig. 1 is explained in details in [10] and [11].

Fig. 2 presents the detailed configuration of the real 4 kV distribution loop studied in the paper, which consists of two 4 kV feeders connected by a re-closer. The loop is energized by two 13 kV/4 kV unit substation transformers. The total load demand of this system is approximately 3 MVA at a power factor of 0.85 and it is shared by about 40 loads.

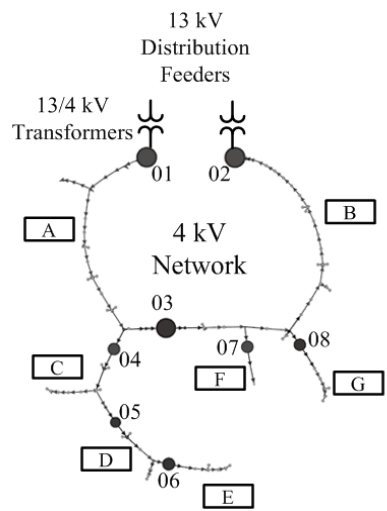

*Fig. 2. Detailed 4 kV distribution system showing protective devices 01 to 08.*

Protective devices 01 and 02 are circuit breakers and device 03 is a re-closer. Those three protective devices open at overcurrent, under-voltage and overvoltage violations, and re-close to recover after the fault. The reclosing prevents unnecessary breaker opening and power outage caused by temporary short circuits. Devices 04, 05, 06, 07 and 08 are fuses which cannot re-close automatically.

There are seven load areas sectionalized by the protective devices, labelled from A to G. Area A is sectionalized by devices 01, 03 and 04; area B is sectionalized by devices 02, 03, 07 and 08; area C is sectionalized by devices 04 and 05; area D is sectionalized by devices 05 and 06; area E is separated by device 06; area F is separated by device 07; and area G is separated by device 08.

The DG used in this study is a synchronous generator which has an active power output of 1.2 MW at a power factor 0.8 (leading). Models for the exciter controller and governor of the generator have been developed. The protection of the D is modelled as in [12].

It is not possible to install DG in areas D and E because the loads in those two areas are two small to consume the power that the DG generates. The back-feeding currents (in steady state) through fuses 05 and 06 would be 103 A and 114 A, respectively, which are larger than the fuse's pickup current of 100 A. Therefore, the legitimate locations for DG are areas A, B, C, F, and G.

The protective device characteristic curves are shown in Fig. 3. Circuit breakers (devices 01 and 02) have identical pickup current rating of 720 A. The pickup current rating of the re-closer is set to be 570 A. Device 04 (fuse) has a pick up current rating of 140 A. The other four fuses (devices 05, 06, 07, and 08) have a pickup of 100 A, which is less than device 04. For the breakers, the tripping time ranges from 0.027 s to 30 s; for the re-closer, the tripping time ranges from 0.16 s to 8 s; for the 140 A fuse, the acting time ranges from 0.01 s to 3.4 s; for the 100 A fuse, the acting time ranges from 0.01 s to 0.64 s.

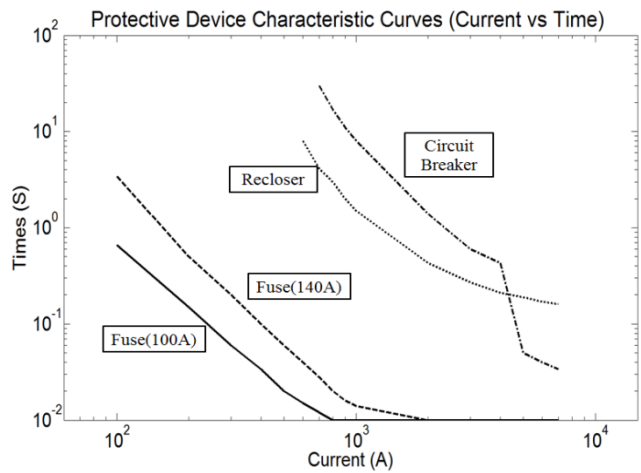

*Fig. 3.* *Protective device characteristic curves*

The re-close time delay of circuit breakers and re-closer is set to 0.4 s, which is 6 cycles slower than the fastest reclose time which is 0.3 s [13] due to possible mechanical time delays. The breaker will wait for 30 s to re-close after the initial opening, 60 s before the second attempt and then 90 s before the third attempt. If the fault still exists, the device will lockout. The re-closer waits for 100 s before the reclose attempt, if it continues to detect the overcurrent, it will open again and stay lockout.

## 3. Case Analysis

*3.1 Single Phase Fault without DG Installed*

A line to ground fault is simulated in area B and at the section next to the feeder breaker 02 which is the very first section of the feeder on the upper right hand side (Fig. 2). The short circuit lasts for 0.3 s from 200 ms o 500 ms; see Fig. 4.

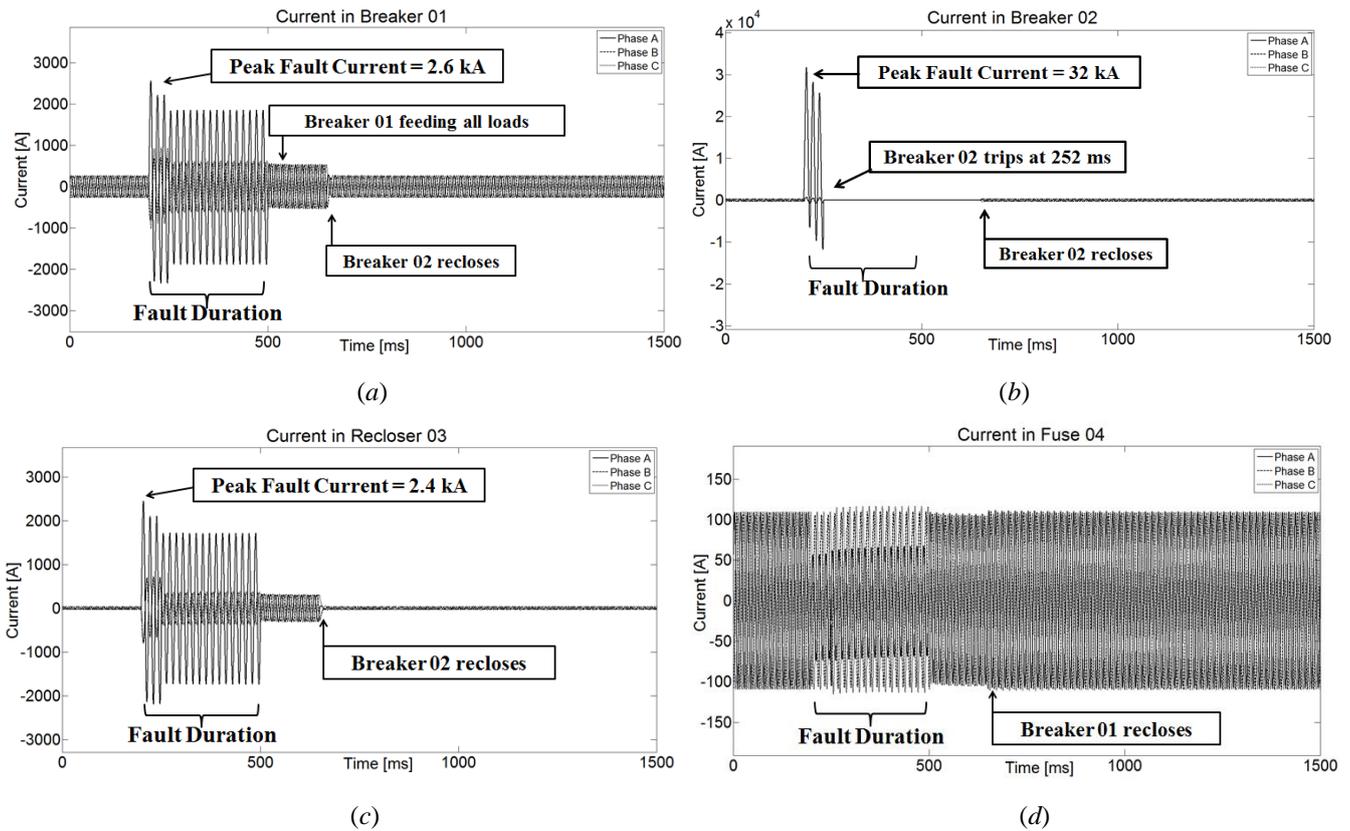

***Fig. 4.*** *Three phase instantaneous currents of recloses from 01 to 04 without DG.*
(a). Three phase instantaneous current of recloses 01
(b). Three phase instantaneous current of recloses 02
(c). Three phase instantaneous current of recloses 03
(d). Three phase instantaneous current of recloses 04

Breaker 02 is the first device to trip on overcurrent from the 13 kV distribution feeders (Fig. 4b) after 3 cycles. The overcurrent peaks at 32 kA. After the single phase fault is cleared, breaker 02 recovers the pre-fault current after the first re-closing attempt.

The highest fault overcurrent of breaker 01 reaches 2.6 kA (Fig. 4a), which does not meet the condition to trip the breaker during this temporary fault. According to the relay characteristic curve (Fig. 3), the circuit breaker takes about one second to open at this current level. During the period when fault is cleared but breaker 02 is still open, breaker 01 is deeding all the loads of the 4 kV network. The current it carries is much higher than its normal current. After 650 ms, when breaker 01 re-closes, breaker 02 recovers its normal current.

Re-closer 03 experiences similar conditions as breaker 01 (Fig. 4c). Its steady state overcurrent during fault is about 2.4 kA. The re-closer does not open during the fault. Between the times when fault is cleared but breaker 01is still open, re-closer 03 is feeding the loads of areas B, G and F. After breaker 02 re-closes, the re-closer current drops back to pre-fault normal level.

During the entire event of the fault (occurrence, clearing, and re-closing), fuse 04 does not experience any overcurrent condition so that it does not blow up. This exactly the way it was intended the system to work (save the fuse). After breaker 01 re-closes, its current level goes back to normal as shown in Fig. 4d.

In this case, every protective device remains closed or reclosed to recover the original customers, there are no customers interrupted in the end. Since the numerator of SAIFI in this case is zero, the SAIFI itself stays zero. Therefore, the reliability is not affected (positively nor negatively) by this temporary event. This is how the system was designed to operate much earlier than DG became popular. As one can see that the system is resilient to temporary faults.

*3.2 Single Phase Fault with DG Installed*

In this case, the temporary line-to-ground fault occurs when the DG is installed. The fault still happens at the section next to feeder breaker 02. The DG is installed at the very last section of area C which is located immediately upstream of fuse 05. This is the case in which there is the longest distance possible between the DG installation and the fault location (Fig. 2) so that the DG has the least fault contribution current. The single line-to-ground fault on phase A lasts for 0.3 s from 200 ms to 500 ms; see Fig. 5.

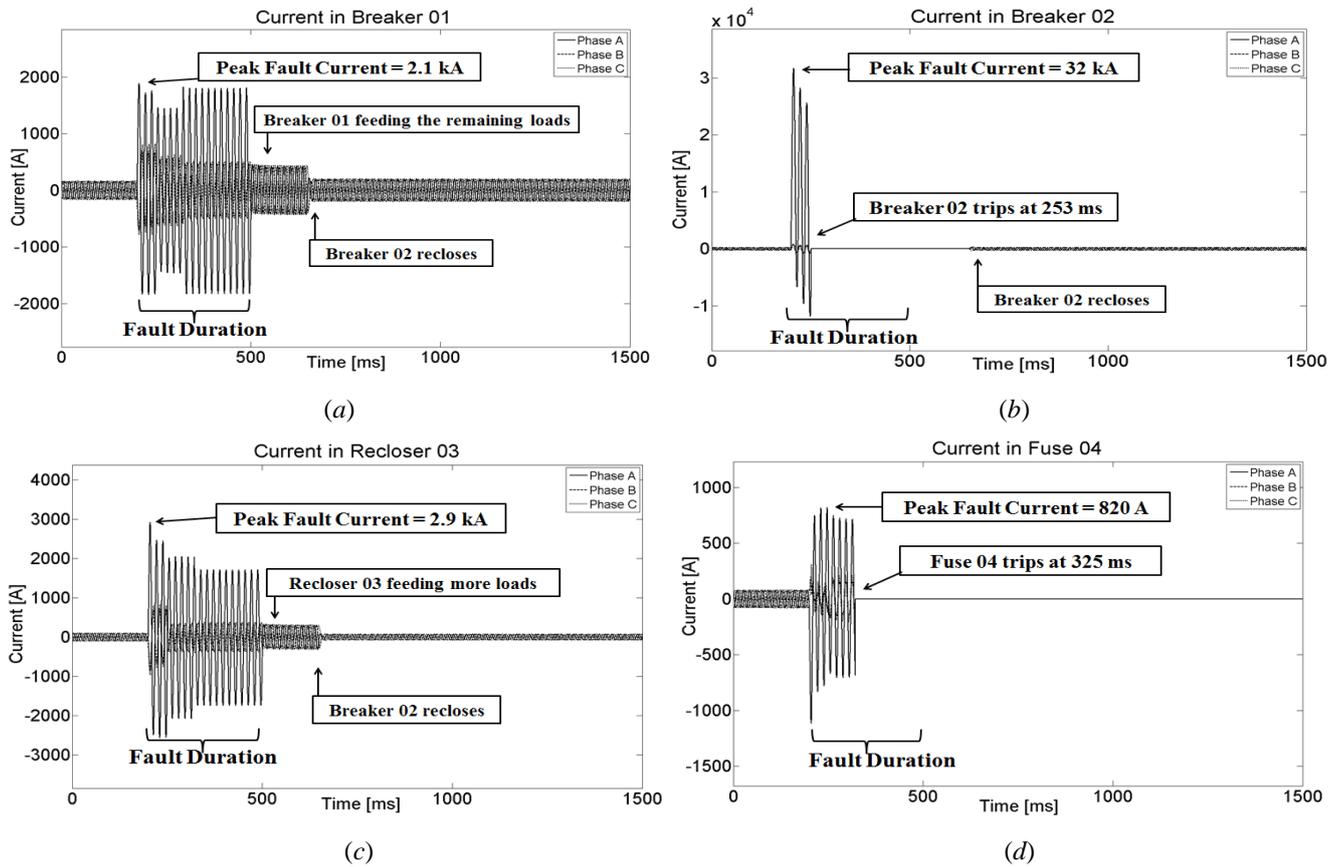

***Fig. 5.*** *Three phase instantaneous currents of recloses from 01 to 04 with DG.*

(a). Three phase instantaneous current of recloses 01

(b). Three phase instantaneous current of recloses 02

(c). Three phase instantaneous current of recloses 03

(d). Three phase instantaneous current of recloses 04

During the entire scenario, the most important device operating is the opening of fuse 04, as shown in Fig. 5d. When the fault happens in area B, there are three contribution paths to the fault, one from breaker 02, the second one from breaker 01 which travels through re-closer 03 and the third one from the DG which flowing through fuss 04 and 03. Fuse 04 opens at 325 ms, roughly 7 cycles after the fault happens. The overcurrent from the synchronous generator reaches 820 A. The fuse current becomes zero and customers located downstream of fuse 04 are lost. The opening of fuse 04 causes long term power outage for customers so that it is the determining factor that negatively affects the reliability index SAIFI.

As shown in Fig. 5b, breaker 02 trips at a very strong 32 kA overcurrent after 3 cycles and re-closes after 400 ms. The post fault current is 260 A which is slightly higher than the 240 A pre-fault current. This is so because the DG was supplying part of the load and now it is disconnected.

Furthermore, the fault overcurrent of breaker 01 reaches 2.1 kA but this overcurrent is not high enough to force the breaker open during the short time span of the temporary fault (Fig. 5a). Between the time intervals when fault is cleared and breaker 02 is still open, breaker 01 is feeding all the loads except the ones located downstream of fuse 04. The pre-fault current of breaker 01 is 160 A, and after breaker 02 re-closes its current drops from 440 A to the new steady state current of 200 A.

It can also be observed from Fig. 5c that re-closer 03 experiences similar overcurrent profiles as breaker 01. The overcurrent peaks at 2.9 kA. That current is not enough for the re-closer to trip during the fault. Between the time intervals when fault is cleared and breaker 02 is still open, re-closer 03 is feeding the loads of area A, since areas C, D, and E are sectionalized by fuse o4 and are isolated by it after the fault. The re-closer's per-fault current is 100 A, after breaker 01 closes, the re-closer's current drops from 310 A to 80 A.

The case presented shows that even the smallest fault current contribution from the DG could trip fuse 04. Considering that the network studied is a relatively large one (3 MVA) with high impedance, the problem reported here where a fuse trips because the installation of DG can happen to any 4 kV loop distribution system in the field.

As shown, when the DG is not present, no customers lose power permanently. Every customer is picked up after the fault is cleared. However, when DG is installed, all the customers located downstream of fuse 04 lose power permanently and they would have to wait for the utility to send crews to restore manually. If this temporary fault is considered the sole incident of the year, the total number of customer interruptions changes from none to a positive number. Therefore, SAIFI is a positive number and the overall reliability of this particular system is reduced by the presence of the DG.

**Table 1** Summary of tripping devices under different DG conditions

| No DG | | DG Installed | | |
|---|---|---|---|---|
| (1) | (2) | (3) | | (4) |
| Fault Location (Area) | Tripping Device | DG Locations Produce Negative Effects (Area) | | Additional Tripping devices |
| A | None | C   F   G | | 04   07   08 |
| B | None | C   F   G | | 04   07   08 |
| C | 04 | F | | 07 |
| D | 05 | A   B   C   F   G | | 04   04   04   04   04 |
| E | 06 | A   B   C   F   G | | 05   05   05   05   05 |
| F | 07 | G | | 08 |
| G | 08 | F | | 07 |

Table 1 presents a summary of the complete analysis of the possible installation location for DG. Table I is divided in four sub-columns: (1) fault location, (2) tripping devices without DG, (3) DG locations which produce negative effects and (4) additional tripping devices with DG installed. The first column shows every possible fault location, from area A to G. The second column shows the corresponding tripping device hen the fault happens in column 1. For faults on areas A and B, there is no opening device, since in both cases the tripping devices are either breaker 01 or breaker 02 which have re-closing functions. During other fault cases, fuses blow and some customers lose power.

Column (3) provides the possible DG locations which produce the negative effect on reliability. Column (4) displays which devices trip because of the presence of DG. For instance if the fault happens in area A and DG is in location C, fuse 04 would be the sole tripping device as no device opens in the fault only case; if the fault occurs in area D and the DG is located in area A, the tripping devices should be fuse 05 accompanied by fuse 04 The additional tripping device makes the reliability index worse than the original.

The statistical summary shown in Table 1 clearly demonstrates the fact that the distribution system reliability. Since the fault overcurrent from synchronous DGs is high, the fuses blow out very quickly. Note that whenever there is a fault in area D/E (downstream of fuse 05/06), the DG installation at any other load area will have a negative impact on reliability index. Those two are the most vulnerable areas.

With 7 possible sections to place faults and 5 sections install DGs, there are 35 possible cases to analyse. Of those, 19 cases cause unfavourable influences on the reliability index, thus there are 54% of the cases can produce worse SAIFI than the original fault cases, as shown in Table 2

**Table 2** Percentage of scenarios with worse reliability index

| | |
|---|---|
| Possible scenarios | 35 |
| Scenarios with worse reliability index | 19 |
| Percentage | 54% |

## 4. Recommendations and solutions

For the utility to prevent the problem to happen, one simple regulatory solution is to never install, or allow any customer to install a DG downstream of a lower rating protective device. Restricting DG installation in area A and B can minimize the chance that negative effects on reliability are produced, as the statistical analysis suggests (see Table 1).

The other solution is to replace the fuses with protective device with more sophisticated settings and functions. For example, a directional relay could be applied to prevent inappropriate coordination when DG is involved. If the fault happens downstream of the relay, the combined fault contribution overcurrent is flowing into the branch towards the fault location, disregarding whether the DG is present or not. Therefore, the relay should act at normal speed; as seen in Figs. 6(a) and (b).

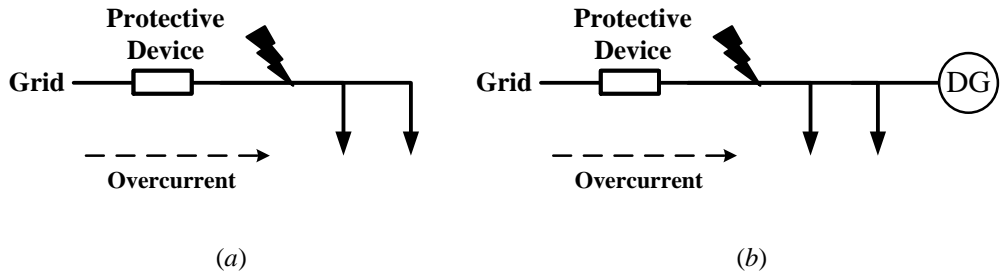

(*a*)            (*b*)

***Fig. 6.*** *Overcurrent at downstream of protective device.*
(a). without DG
(b). with DG

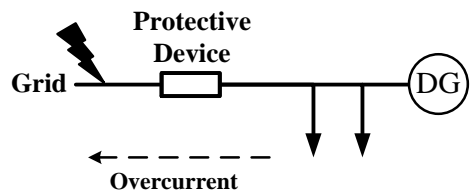

***Fig. 7.*** *Reverse overcurrent at upstream of protective device.*

On the other hand, if the DG is present and the fault happens upstream of the device as shown in Fig. 7, the overcurrent wold be flowing out of the branch so that the relay would not operate. The DG's protection relay needs to trip to stop overcurrent contribution and stop the generator.

With the help of modern microprocessor relays, the variation of relay setting can be achieved very easily so that the utility can modify the rating anytime there is a change of the local power demand. The only problem preventing this solution is the economic concern. More advanced relays require extra meters and additional field tests during installation. The supplementary equipment expenditures and labour costs hinder the utility to execute such a replacement. The other alternative is to have the customer installing the DG to pay for the change of protection devices. This frequently will kill the project.

## 5. Conclusions

Time domain dynamic simulations show clearly how the protective devices are affected by the fault contribution current from the DG. The comparison between two scenarios proves that the inappropriate installation of DG will inevitably cause negative effect on SAIFI during a temporary fault.

The statistical analysis shows that there are a significant number of scenarios that can introduce negative effects on reliability. The results also demonstrate the necessity of detailed dynamic modelling of distribution systems involving accurate protective relay coordination.